\documentclass[aps%
,prl%
,showpacs%
,twocolumn%
,superscriptaddress%
,floats%
,amssymb%
]{revtex4}

\usepackage{graphicx}
\usepackage{amsmath}
\usepackage{amssymb}
\bibstyle{apsrev.bib}

\newcommand{\beq}{\begin{equation}}
\newcommand{\eeq}{\end{equation}}

\newcommand{\comment}[1]{}

\begin{document}
\title{Stiffness of the Edwards-Anderson Model in all Dimensions} 
\author{Stefan Boettcher
} 

\affiliation{Physics Department, Emory University, Atlanta, Georgia
30322, USA} 
\begin{abstract}
A comprehensive description in all dimensions is
provided for the scaling exponent $y$ of low-energy
excitations in the Ising spin glass introduced by Edwards and
Anderson. A combination of extensive numerical as well as theoretical
results suggest that its lower critical dimension is {\it exactly}
$d_l=5/2$. Such a result would be an essential feature of any complete
model of low-temperature spin glass order and imposes a constraint
that may help to distinguish between theories. 
\end{abstract} 


\maketitle
Imagining physical systems in non-integer dimensions, such as through
an $\epsilon$-expansion near the upper~\cite{Wilson72} or lower
critical dimension~\cite{Parisi}, has provided many important results
for the understanding of the physics in realistic dimensions. Often,
peculiarities found in unphysical dimensions may impact real-world
physics and enhance our understanding~\cite{Bender92}. Here, we will
explore the variation in dimension of the low-temperature behavior in
the Edwards-Anderson (EA) spin glass model~\cite{Edwards75}.

A quantity of fundamental importance for the modeling of amorphous
magnetic materials through spin glasses~\cite{F+H} is the
``stiffness'' exponent $y$~\cite{Southern77,Bray84}. As Hooke's law
describes the response in increasing elastic energy imparted to a
system for increasing displacement $L$ from its equilibrium position,
the stiffness of a spin configuration describes the typical rise in
magnetic energy $\Delta E$ due to an induced defect-interface of size
$L$. But unlike uniform systems with a convex potential energy
function over its configuration space (say, a parabola for the sole
variable in Hooke's law), an amorphous many-body system exhibits a
function more reminiscent of a high-dimensional mountain
landscape~\cite{Frauenfelder96}. Any defect-induced displacement of
size $L$ in such a complicated energy landscape may move a system
through many ups-and-downs in energy $\Delta E$. Averaging over many
incarnations of such a system results in a typical energy scale
\begin{eqnarray}
<\vert\Delta E\vert>\sim L^y\quad(L\to\infty).
\label{yeq}
\end{eqnarray}
The importance of this exponent for small excitations in disordered
spin systems has been discussed in many
contexts~\cite{Southern77,Bray84,Fisher86,Krzakala00,Palassini00,Bouchaud03,F+H}.

Spin systems with $y>0$ provide resistance (``stiffness'') against the
spontaneous formation of defects at sufficiently low temperatures $T$;
an indication that a phase transition $T_c>0$ to an ordered state
exists. For instance, in an Ising ferromagnet, the energy $\Delta E$
is always proportional to the size of the interface, i.~e. $y=d-1$,
consistent with the fact that $T_c>0$ only when $d>1$. When $y\leq0$,
a system is unstable (such as the $d=1$ ferromagnet) to spontaneous
fluctuations which proliferate, preventing any ordered state. Thus,
determining the ``lower critical dimension'' $d_l$, where $y_{d_l}=0$,
is of significant
importance~\cite{Southern77,McMillan84b,Fisher88,Maucourt98}, and
understanding the mechanism leading to $d_l$, however un-natural,
provides definite clues to the origin of order~\cite{Parisi}. For
instance, in homogeneous systems with a continuous
symmetry~\cite{Goldenfeld}, such as the Heisenberg ferromagnet, the
possibility of ``soft modes'' (Goldstone bosons) perpendicular to the
direction of magnetization have proved to weaken order further, since
defect-interfaces can broaden over the entire length $L$ of the system
(spin wave) to reduce the energy by a factor $\sim1/L$. This effect
manifests itself in a reduced stiffness, $y=d-2$, and an increased
$d_l=2$.

The addition of quenched disorder drastically complicates the nature
of the ordered state~\cite{Imry75}, since fluctuating interfaces can
attain highly irregular geometries to take advantage of
heterogeneities. In this case, $y$, and possibly $d_l$, may become
anomalous. For instance, it was found that adding even a small random
field (of zero mean) to an ordinary Ising ferromagnet destabilizes
order at least in $d=2$, and considerable insight into the ordered
state was gained in proving that
$d_l=2$~\cite{Andelman84,Imbrie84,Bricmont87}.

For spin glasses, even a numerical determination of $y$ is highly
nontrivial when $d>2$~\cite{Barahona82}, requiring repeated solution
of some of the hardest-known combinatorial optimization
problems~\cite{Dagstuhl04}. Thus, simulations have been rather limited
in system size $L$ and yielded widely varying results with
non-overlapping errors, even for $d=3$~\cite{Boettcher04c}. The
consensus from these studies is that $2\leq
d_l\leq3$~\cite{Amoruso03}, where $y_d$ may behave in a discontinuous
manner~\cite{Hartmann01}, such that $d_l$ could be integer as is the
case for other statistical models.  On the other extreme, above the
``upper critical dimension,'' well-known to be $d_u=6$~\cite{F+H},
where fluctuations become irrelevant, numerical results can be
compared with analytical predictions for critical exponents from
mean-field theory~\cite{Aspelmeier03}.

In this Letter, we consider an analytic continuation of $y=y_d$ to
real dimensions $d\geq0$, based on extensive numerical simulations for
$y_d$ in integer dimensions~\cite{Boettcher04b,Boettcher04c}. That data
suggest that $y_d$ is varying smoothly across dimensions and that a
straightforward low-order fit would provide a very accurate
description. The validity and accuracy of this fit is gauged by how
well it extrapolates to an exactly known feature, e.~g. $y_1=-1$. The
fit has a zero at $d_l\approx2.4986$ and yields
$y_{5/2}\approx0.0008$; strong evidence that $d_l=5/2$. This value can
be corroborated by a similar but independent fit of the existing data
for $T_g$~\cite{Ballesteros00,Bernardi97,Klein91}. A speculative
calculation based on mean-field arguments~\cite{Franz94}, recently put 
on a more rigorous basis~\cite{Franz05}, has predicted
the disappearance of replica symmetry breaking (RSB)~\cite{MPV}, taken
to be a characteristic of low-$T$ glassy order, for dimensions
$d<d_l=5/2$ exactly. Our results here lend significant support to the
low-$T$ structure of spin glasses advanced in Refs.~\cite{Franz94,Franz05}.

Applying a new optimization heuristic~\cite{Boettcher01a} to spin
glasses on {\it bond-diluted} lattices~\cite{Boettcher04c} with up to
$10^6$ spins allowed us to present values for $y_d$ in $d=3,\ldots,7$
that are improved or are entirely unprecedented: $y_3=0.24(1)$,
$y_4=0.61(1)$, $y_5=0.88(5)$, $y_6=1.1(1)$, and $y_7=1.24(5)$. The
precision of these values originates from extensive numerical
simulations~\cite{Boettcher04b} over a large number of systems sizes and
bond fractions, yielding 30 or more data points significant for
scaling in each dimension, with each point resulting from an average
over $10^4-10^6$ instances.

We can supplement this list of values for $y_d$ by those for $d=1$ and
2, assuming a bond distribution that in addition to being symmetric
and of unit width also has to be continuous at the origin. Only for
$y_d>0$ is its value independent of details of the bond
distribution~\cite{Bouchaud03}, where the energy scale in
Eq.~(\ref{yeq}) diverges. For $d\leq2$, discrete $\pm J$ bonds appear
to lead to trivial scaling,
$y_{d\leq2}\equiv0$~\cite{Hartmann01,Amoruso03}, whereas continuous
bonds generally provide nontrivial results~\cite{Amoruso03}. For
instance, in a $d=1$ spin chain of length $L$, any defect at $T=0$
affects only the absolute weakest bond. Hence, for $\pm J$ bonds, or
any other distribution bounded away from the origin, the absolute
weakest bond is a non-zero constant for large $L$. For a distribution
continuous at the origin, the absolute smallest of $L$ bonds scales as
$\sim1/L$, thus $y_1=-1$ exactly. The value for $y_2$ has been
considered repeatedly over the
years~\cite{Southern77,Bray84,Rieger96,Hartmann02,BoHa}, and has been
determined with some consistency to be $y_2=-0.282(2)$ for continuous
bonds. The accuracy for $y_2$ is expected to be significantly better
than for any $y_d$ with $d>2$, since finding exact ground states for
$d=2$ spin glasses is accomplished with polynomial time algorithms,
facilitating larger sizes and better statistics.

\begin{figure}
\vskip 2.2in \includegraphics{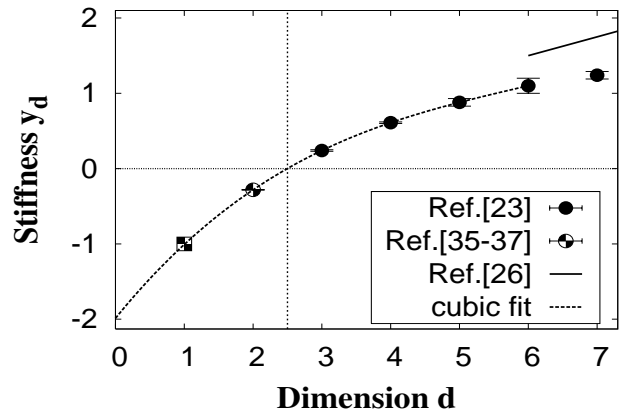}
\caption{Plot of the spin glass stiffness exponent $y_d$ as a function
of dimension $d$. Shown are the data for hypercubic lattices from
Refs.~\protect\cite{Boettcher04c,Rieger96}, a weighted cubic fit to
the $y_d$ data on $2\leq d\leq6$, and the RSB prediction from
Ref.~\protect\cite{Aspelmeier03} in Eq.~(\protect\ref{stiffeq}). The fit
predicts to $0.1\%$ that the lower critical dimension defined through
$y_{d_l}=0$ is $d_l=5/2$ (vertical line). It reproduces the exact
result, $y_1=-1$ (checkered square), to $0.8\%$ and suggests
$y_0=-2$. Above the upper critical dimension, $d_u=6$, the 
data significantly deviates from Eq.~(\protect\ref{stiffeq}) (solid
line).}
\label{SKstiff}
\end{figure}

{}From these values, a clear picture for the stiffness exponent
emerges. As Fig.~\ref{SKstiff} shows, the exponent appears to vary
smoothly with dimension $d$, at least between $1\leq d\leq6$. (It can
be expected that $y_d$ would develop a cusp at the upper critical
dimension $d_u=6$.) Hence, we expect that a simple polynomial fit of
the numerical data may provide an accurate interpolation for $y_d$. In
light of the much reduced uncertainty in the value of $y_3$ compared
to earlier results, which ranged between $0.19(1)$ to
$0.27$~\cite{Boettcher04c}, such a fit should provide an accurate
prediction for $d_l$, i. e. the zero of $y_d$. To fit the data, we
proceed in the following way: We fit {\it only} the numerical data for
$2\leq d\leq6$, weighted by their error bars; the quality of the fit
is judged on the basis of how well the exactly known data point
$y_1=-1$ is reproduced. We find that the optimal fit is of cubic
order: a square fit does not even have enough variability to match all
the included data well, let alone $y_1$; fits of higher order become
unstable since the number of fitting parameters approaches the number
of data points. The resulting cubic fit,
\begin{eqnarray}
y_d\approx-1.988+1.125d-0.1533d^2+0.0086d^3,
\label{ydfiteq}
\end{eqnarray}
has a number of desirable properties. In particular, the coefficients
for increasing powers of $d$ rapidly decrease, testifying to the
stability of the fit. Consequently, Eq.~(\ref{ydfiteq}) {\it predicts}
the exactly-known result, $y_1=-1$, to less than $0.8\%$ and further
extrapolates to $y_0=-2$ (within $0.7\%$). The fit has a zero at
$d_l\approx2.4986$ and yields $y_{5/2}\approx0.0008$; suggesting that
$d_l=5/2$.

In fact, a replica theory calculation on a lattice in arbitrary
dimensions~\cite{Franz94} yields $d_l=5/2$ exactly. It uses a
Landau-like expansion in the neighborhood of $T_g$ for the order
parameter function for two coupled replicas. Both replicas are held at
an identical equilibrium state on one open boundary, and held at two
distinct states on the opposite boundary. While the ``defect'' thus
created leads to a different exponent, say, $a_d\not\equiv y_d$, their
zero's should coincide. The defect free energy $\Delta F$ signals that
the distinction between overlaps becomes irrelevant exactly in
$d=5/2$, where replica symmetry gets reestablished throughout. The
ad-hoc free-energy extremization in Ref.~\cite{Franz94} has recently
been put on a more rigorous basis~\cite{Franz05}, using the inverse
range as an expansion parameter near a saddle point of a lattice spin
glass with variable interaction range.  It is argued that this $\Delta
F$ behaves similar to that in ordinary ordered systems with a
continuous symmetry, such as the Heisenberg ferromagnet, in which the
Goldstone mode undermines order already in
$d_l=2$~\cite{Goldenfeld}. Yet, in the Ising spin glass order is
further weakened by the bond disorder without any ferromagnetic bias,
making any contribution to the defect energy sub-additive
($\langle\Delta F\rangle=0$ but $\langle|\Delta F|\rangle>0$), hence
$a_d=d-2-\frac{1}{2}$. The replica calculation in
Refs.~\cite{Franz94,Franz05} is presently the only realistic theory
with an analytic prediction for $d_l$. Considering the simplicity of a
$d_l=5/2$, one may wonder how essential the replica argument is.

We can put these predictions further into context by comparing with
the Migdal-Kadanoff approximation (MK)~\cite{Migdal76,Kadanoff76},
which has proved to resemble many features of low-dimensional spin
glass quite well~\cite{BoCo}. In MK, a $d$-dimensional lattice is
represented by a hierarchical graph where each bond of range $s\times
L$ between two spins consists recursively of $b$ parallel series of
$s$ bonds of range $L$, leading
to~\cite{Southern77,Amoruso03,Bouchaud03}
\begin{eqnarray}
d=1+\frac{\ln b}{\ln s}.
\label{MKeq}
\end{eqnarray}
Recent simulations of MK in non-integer $d$ (using $b=3$ and
continuous $s$) have come surprisingly close to the same
result~\cite{Amoruso03}, $d_l\approx2.5$. Already
Ref.~\cite{Southern77} found analytically for MK, within a Gaussian
approximation, that
\begin{eqnarray}
y_d^{MK}\approx\frac{d-1}{2}+\frac{\ln\left(1-\frac{2}{\pi}\right)}{2\ln2},
\label{SYeq}
\end{eqnarray}
which has been improved to a $d\to\infty$ expansion of $y_d$ in
Ref.~\cite{Bouchaud03}. Clearly, Eq.~(\ref{SYeq}) fails at $d=1$, and
its linear form misses the essential features of the data in
Fig.~\ref{SKstiff}, but it provides a decent estimate for $d=2$ and 3,
and predicts $d_l^{MK}\approx2.46$. (Similarly, a suggestion from
Ref.~\cite{Cieplak90} that $(d-1)/2-y=const$ is ruled out by the data.)

Yet, MK predictions are generally ambiguous, since various
combinations of $b$ and $s$ in Eq.~(\ref{MKeq}) can represent the same
$d$. For instance, for $s\gg1$ each series of bonds at $T=0$ can be
replaced by the (absolute) smallest bond (like the $d=1$ spin chain
above). Drawn from a Gaussian distribution of zero mean and unit
variance, the equivalent bond replacing each series is of typical size
$\sim1/s$ and has a random sign. Putting $b$ such {\it iid} bonds in
parallel thus yields another effective bond of Gaussian distribution
with zero mean and width $\sim\sqrt{b}/s$. Hence, for an interfacial
energy $\Delta E$ to be scale invariant [i. e. $y=0$ in
Eq.~(\ref{yeq})] requires $b=s^2$ for $b,s\to\infty$, implying
$d_l^{MK}=3$ according to Eq.~(\ref{MKeq}). This result is at odds
with computations for $b=3$~\cite{Amoruso03} and variable
$s$. Furthermore, using $s=2$ and an analytic continuation for
$b\to1/2$, one can show~\cite{unpublished} that $y_0^{MK}=-1$; quite
inconsistent with Fig.~\ref{SKstiff}.

\begin{figure}
\vskip 2.2in \includegraphics{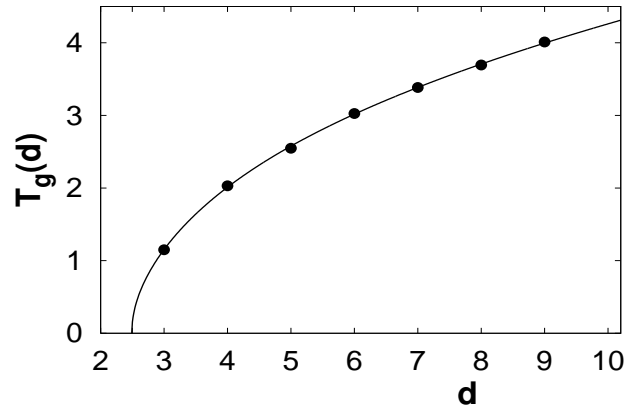}
\caption{Plot of the glass transition temperature $T_g(d)$ as a
  function of dimension $d$. Data points are drawn from various
  studies~\protect\cite{Bernardi97,Marinari98,Ballesteros00,Klein91,Daboul04}. The solid line represents the fit in
  Eq.~(\protect\ref{Tgeq}). That fit captures the known asymptotic behavior 
$T_g(d)\sim\sqrt{2d}$ for $d\to\infty$ and predicts a zero at $d\approx2.491$.}
\label{Tgfit}
\end{figure}

As an independent test for the picture emerging from
Fig.~\ref{SKstiff}, we study the values for the glass-transition
temperatures as a function of dimension, $T_g(d)$, for $d\geq3$. To
this end, we have gathered the latest values for $T_g(d)$ (using $\pm
J$ bonds) from the literature, based on
simulations~\cite{Bernardi97,Marinari98,Ballesteros00} or high-$T$
series expansions~\cite{Klein91,Daboul04}. We found
$T_g(3)\approx1.15$, $T_g(4)\approx2.03$, $T_g(5)\approx2.548$,
$T_g(6)\approx3.026$, $T_g(7)\approx3.385$, $T_g(8)\approx3.694$, and
$T_g(9)\approx4.011$. Determining $d_l$ by fitting $T_g(d)$ is
somewhat more complicated, since it vanishes for all $d<d_l$ (unlike
$y_d$) and Ref.~\cite{Fisher88} argues that $T_g(d)\sim\sqrt{d-d_l}$
for $d\to d_l^+$. On the other hand, for the Sherrington-Kirkpatrick (SK)
model $T_g^{SK}=J^{SK}$~\cite{Sherrington75} and
$J\sim J^{SK}/\sqrt{2d}$, so $T_g(d)\sim\sqrt{2d}$ for
$d\to\infty$ and $J=1$. Satisfying these constraints, we fit
\begin{eqnarray}
T_g(d)\approx\sqrt{2\left(d-2.491\right)}\left(1+\frac{1.237}{d}-\frac{2.425}{d^2}\right),
\label{Tgeq}
\end{eqnarray}
which is plotted in Fig.~\ref{Tgfit}. Again, incorporating more terms
in the fit leads to drastic oscillations in the coefficients. While
the predicted result, $d_l\approx2.491$, is not quite as accurate or
stable as above, it is quite consistent.\footnote{A similar fit has been
attempted already in 
G. Parisi,  P. Ranieri,  F. Ricci-Tersenghi, and J. J. Ruiz-Lorenzo,
J. Phys. A: Math. Gen. {\bf 30}, 7115 (1997).}

Unlike the consistency between mean-field arguments and numerics for
$d_l$, we can observe a large discrepancy for $d\geq d_u=6$. An RSB
calculations from Ref.~\cite{Aspelmeier03} around the mean-field limit
($d\to\infty$)~\cite{Sherrington75} predicts
\begin{eqnarray}
y_d=d\left(1-\rho\right)\approx\frac{d}{4}\qquad(d\geq d_u=6),
\label{stiffeq}
\end{eqnarray}
where the exponent $\rho$, predicted~\cite{Aspelmeier03,Bouchaud03} and 
measured~\cite{Bouchaud03,Palassini03,EOSK} to be $\approx3/4$, 
describes the width of the
distribution of ground state energies in the SK model.  As
Fig.~\ref{SKstiff} shows, the finite-dimensional results substantially
disagree on this point with replica theory. A plausible interpolation
may be $y_d\approx(d-2)/4$, which approaches Eq.~(\ref{stiffeq}) for
$d\to\infty$.

In conclusion, we have provided a series of {\it independent, mutually
consistent} arguments for a lower critical dimension of $d_l=5/2$ for
the Edwards-Anderson spin glass. This self-consistency has a number of
important implications. For one, it validates the numerical values for
$y_d$ for integer dimensions as quoted
above~\cite{Boettcher04b,Boettcher04c}; any change beyond the error bars of
$y_3$ in particular would significantly distort the obtained
$d_l$. The accuracy of those values, in particular $y_6$ and $y_7$,
makes the comparison with the mean-field prediction for $d\geq6$
relevant, allowing for a rare {\it direct} comparison between lattices
and mean-field predictions across the dimensions. In turn, the
apparent square-root singularity near $d_l$ validates the arguments
from Ref.~\cite{Fisher88} for the behavior of $T_g$. Most importantly,
the consistency of these numerical results with the analytical
treatment in Refs.~\cite{Franz94,Franz05} lend credibility to its
assumptions and implications.  That theory assumes replica symmetry
breaking to exist below $T_g$ for finite-dimensional lattices, which
seems incompatible~\cite{Newman96} with the droplet scaling picture
developed in Refs.~\cite{Bray84,Fisher86}. It further derives that the
treatment is unaffected by the existence of an external field, hence
predicting an de Almeida-Thouless line, in contrast with some recent
experimental~\cite{Jonsson05} and numerical
findings~\cite{Young04}. Our findings here suggest a careful
reexamination of those ideas.

Many thanks to S. Franz and M. A. Moore for their valuable comments on the
manuscript. This work has been supported by grant 0312510 from the Division of
Materials Research at the National Science Foundation and by the Emory
University Research Council.

\comment{
These commands specify that your bibliography should be created from entries in the file bibdata.bib. There are several bibliography styles that can be used with the eethesis document class; the style theunsrt.bst is the style used in the examples in this manual. The entries in this style are patterned after those in IEEE Transactions on Automatic Control. It lists the sources in the order they were cited. There is also a generic ieeetr.bst, which formats sources similar to many IEEE publications. If neither of these styles is suitable for your department, you might consider acm.bst or siam.bst which format your bibliography in the style of ACM and SIAM publications. Check the /usr/local/teTeX/local.texmf/bibtex/bst directory on the machine that you use to see if there are any other .bst files you can use.}
\bibliographystyle{apsrev}
\bibliography{lowDprl2}

\end{document}